 \documentclass[preprint]{aastex}

\newcommand{\kms}{km s$^{-1}$}

\slugcomment{To appear in the Astrophysical Journal}

\shorttitle{Carbon-Chain Molecules in L1521F and IRAM 04191+1522} 
\shortauthors{Takakuwa, Ohashi, \& Aikawa}

\begin{document}


\title{Carbon-Chain and Organic Molecules\\
around Very Low-Luminosity Protostellar Objects of\\
L1521F-IRS and IRAM 04191+1522}

\author{Shigehisa Takakuwa\altaffilmark{1} and Nagayoshi Ohashi}
\affil{Academia Sinica Institute of Astronomy and Astrophysics,\\
P.O. Box 23-141, Taipei 10617, Taiwan} 

\and

\author{Yuri Aikawa}
\affil{Department of Earth and Planetary Sciences, Kobe University,\\
Kobe 657-8501, Japan}

\altaffiltext{1}{e-mail: takakuwa@asiaa.sinica.edu.tw}

\begin{abstract}
We have observed dense gas around the Very Low-Luminosity Objects (VeLLOs)
L1521F-IRS and IRAM 04191+1522 in carbon-chain and organic
molecular lines with the Nobeyama 45 m telescope. Towards L1521F-IRS,
carbon-chain lines of CH$_{3}$CCH (5$_{0}$--4$_{0}$),
C$_{4}$H ($\frac{17}{2}$--$\frac{15}{2}$), and C$_{3}$H$_{2}$ (2$_{12}$--1$_{01}$)
are 1.5 - 3.5 times stronger than those towards IRAM 04191+1522,
and the abundances of the carbon-chain molecules towards L1521F-IRS are 2 to 5 times higher
than those towards IRAM 04191+1522.
Mapping observations of these carbon-chain molecular lines
show that in L1521F the peak positions of
these carbon-chain molecular lines are different from each other and
there is no emission peak towards the VeLLO position,
while in IRAM 04191+1522 these carbon-chain lines are as weak as the detection limits
except for the C$_{3}$H$_{2}$ line.
The observed chemical differentiation
between L1521F and IRAM 04191+1522 suggests that the evolutionary stage of L1521F-IRS
is younger than that of IRAM 04191+1522, consistent with the extent of the associated
outflows seen in the $^{13}$CO (1--0) line.
The non-detection of the organic molecular lines of
CH$_{3}$OH (6$_{-2}$--7$_{-1}$ E) and CH$_{3}$CN (6$_{0}$--5$_{0}$) implies that the warm ($\sim$ 100 K)
molecular-desorbing region heated by the central protostar is smaller than
$\sim$ 100 AU towards L1521F-IRS and IRAM 04191+1522, suggesting the young age of these VeLLOs.
We propose that the chemical
status of surrounding dense gas can be used to trace the evolutionary stages of VeLLOs.
\end{abstract}


\section{Introduction}

Stars form in dense ($>$ 10$^{5}$ cm$^{-3}$) molecular-cloud cores.
The IRAS survey of dense cores has revealed that approximately half of dense cores
are associated with bright ($>$ 0.1 $L_{\odot}$) infrared
point sources, and the dense cores associated with the IRAS sources
have been considered as dense cores
with the onset of the star formation \cite{bei86,mye87,lee99}.
Recent high-sensitivity mid- and far-infrared
observations with the Spitzer Space Telescope ($\sim$ 25 times higher sensitivity than with IRAS),
however, have found infrared point sources
in dense cores previously thought to be starless.
The first such ``starless'' core is L1014, where Young et al. (2004)
have discovered a Spitzer source (L1014-IRS) with a faint internal luminosity
($L_{int}$ $\sim$ 0.09 $L_{\odot}$).
The internal (star+disk) luminosity, $L_{int}$, is defined as $L_{bol}$ -- $L_{isrf}$,
where $L_{isrf}$, typically $\sim$ 0.2 $L_{\odot}$, is 
the luminosity due to the interstellar radiation field.
Molecular-line observations have found plenty of dense molecular gas
associated with L1014-IRS, which could accrete onto and further raise the mass of the central object \cite{cr05b}.
A compact ($\sim$ 500 AU) molecular outflow has also been found with the SMA in L1014-IRS \cite{bou05}.

Subsequent surveys of low-mass star-forming regions with Spitzer have discovered more
infrared sources in dense cores previously thought to be starless
\cite{bou06,dif06,dun06,dun08,ter09}. Those Spitzer sources with
rising or flat SEDs between the longest
detected Spitzer IRAC wavelength and MIPS 24 $\micron$ as well as between 24 and 70 $\micron$,
and $L_{int}$ below or equal to 0.1 $L_{\odot}$, are dubbed
``Very Low Luminosity Objects'' or VeLLOs \cite{dun08}.
%
So far, $\sim$ 15 VeLLOs have been identified out of 67 ``starless cores'', including L1014-IRS, mentioned above,
and hence $\sim$ 80 $\%$ of these ``starless'' cores remain starless down to
the sensitivity limit of $L_{int}$ $>$ 4 $\times$ 10$^{-3}$ ($d$/140 pc)$^2$ \cite{dun08}.
%
%
%
%
%
%
Because of the characteristics of their SEDs and the very low 
internal luminosity,
those VeLLOs could represent the very early phase of star formation with
a lower central stellar mass and/or mass accretion rate
than those of typical Class 0 sources.
The detection of faint $K_{s}$ emission coincident with L1014-IRS, however,
suggests that L1014-IRS may not be a young protostellar source, although
it is difficult to tell whether part of the $K_{s}$ emission directly
arises from the central star or entirely from
the reflection nebula tracing the envelope cavity presumably evacuated
by an outflow \cite{hua06}.

L1521F is one of ``starless'' cores in Taurus, where the Spitzer survey
has revealed the presence of
a VeLLO, L1521F-IRS ($L_{int}$ $\sim$ 0.06 $L_{\odot}$) \cite{bou06,ter09}.
L1521F-IRS was directly detected at
mid-infrared wavelengths (24 and 70 $\mu$m),
while at shorter infrared wavelengths ($<$ 5 $\mu$m) only the scattered light
was detected.
The scattered light shows a bipolar nebulosity
oriented along the east to west direction, which probably traces an outflow cavity.
The 1.3 mm and 850 $\mu$m continuum maps in L1521F show a centrally-peaked
dusty condensation with a mass of 1.5 $M_{\odot}$ around L1521F-IRS \cite{cra04,shi04}.
The position of the Spitzer source is approximately (within $\sim$ 5$\arcsec$)
consistent with the peak position of the dust-continuum emission \cite{bou06}.
In the central 20$\arcsec$ region the molecular hydrogen density ($n_{\rm H_{2}}$)
is estimated to be $\sim$ 10$^{6}$ cm$^{-3}$, and CO molecules are depleted by a factor of 15
from the nominal CO abundance of 9.5 $\times$ 10$^{-5}$ \cite{cra04}.
BIMA observations of L1521F in the CCS ($J_N$ = 3$_{2}$--2$_{1}$)
and N$_{2}$H$^{+}$ (1--0) lines have shown that the N$_{2}$H$^{+}$ emission
correlates well with the dust-continuum emission while the CCS emission
traces the outer region surrounding the N$_{2}$H$^{+}$ emission \cite{shi04}.
Onishi et al. (1999) have studied L1521F (MC 27 in their paper)
in millimeter and submillimeter HCO$^{+}$ ($J$ = 4--3, 3--2) and
H$^{13}$CO$^{+}$ ($J$ = 4--3, 3--2, 1--0) lines
and have found a highly condensed gas structure and a signature of infalling motion
with an infalling velocity of 0.2 - 0.3 km s$^{-1}$ at
2000 - 3000 AU. They have also found that only the central position of
the HCO$^{+}$ ($J$=3--2) line shows winglike components
(5.0 - 5.8 km s$^{-1}$ and 7.4 - 8.6 km s$^{-1}$), suggesting the compactness
of the associated molecular outflow. From these results, they suggested, before
the discovery of the VeLLO, that L1521F is on the verge of star formation.

IRAM 04191+1522 (hereafter IRAM 04191)
has previously been identified as a Class 0 protostar
with its very low T$_{bol}$ of $\sim$ 18 K,
located in the southern part of the Taurus molecular cloud \cite{an99a,an99b}.
Spitzer observations have revealed that IRAM 04191
is another VeLLO with $L_{int}$ $\sim$ 0.08 $L_{\odot}$
\cite{dun06}. IRAM 04191 was directly detected as an infrared point source
at all six wavelengths observed by Spitzer
(3.6, 4.5, 5.8, 8.0, 24, and 70 $\mu$m), associated with a nebulosity at wavelengths
shorter than 24 $\mu$m extending to the south.
The Spitzer source position agrees within 0.2$\arcsec$
with the source position given by Belloche et al. (2002) based
on the 227 GHz continuum emission detected by the IRAM PdBI interferometer.
IRAM 04191 is associated with a well-developed ($>$ 10000 AU) molecular outflow
extending from northeast (red) to southwest (blue),
which appears to be similar to outflows from more luminous protostars
\cite{an99a,lee02,dun06}.
The southern infrared nebulosity appears to be consistent with the
blueshifted molecular outflow \cite{dun06}.
Millimeter molecular-line observations of IRAM 04191 show
a 10000-AU scale rotating molecular envelope around IRAM 04191 \cite{bel02,tak03,lee05}.
Comparison between the 1.3 mm dust-continuum and N$_{2}$H$^{+}$ (1--0) observations of IRAM 04191
shows that N$_{2}$H$^{+}$ depletion by a factor of $\sim$ 4
is present at the center of the envelope
($r$ $<$ 1600 AU, $n_{\rm H_{2}}$ $>$ 5 $\times$ 10$^{5}$ cm$^{-3}$),
whereas at the outer part the N$_{2}$H$^{+}$ emission correlates
with the dust-continuum emission \cite{bel04}.

It is still not clear, however, how to track the evolutionary
stage and the differences among these VeLLOs.
The chemical status of dense gas around the VeLLOs could be a key,
since chemistry in dense cores is non-equilibrium and
time-dependent (e.g. Leung et al. 1984; Herbst \& Leung 1989).
Carbon-chain molecules are of particular interest.
Since carbon-chain molecules are produced through hydrogenation
of C$^{+}$ and C, and carbon in the form of C$^{+}$ and C is only present at the
early ($\leq$ 10$^{5}$ yr) evolutionary stage of the chemical evolution,
they are likely to trace the early evolutionary stage of star formation.
Observations of organic molecules such as CH$_{3}$OH and CH$_{3}$CN
towards VeLLOs are also intriguing.
At the ``real'' starless phase the low temperatures at the
central regions result in the depletion of most molecules onto dust grains.
Once a protostellar source forms at the center of the dense core,
heating from the protostar desorbs molecules from grain surfaces,
and the desorption region extends as the protostellar evolution proceeds
\cite{sch04,aik08}.
Since organic molecules are thought to be formed on grain surfaces
\cite{aik08}
(for example CH$_{3}$OH is formed through hydrogenation of CO
on grain surfaces), the amount of organic molecules in the gas phase towards VeLLOs could be an independent
measure of the evolutionary stage.

In order to study the chemical status of the molecular material surrounding VeLLOs
and to investigate the evolution of the earliest protostellar formation,
we have conducted observations
of L1521F and IRAM 04191 in several carbon-chain and organic molecular lines
with the Nobeyama 45-m telescope.
Our observations of the two representative
VeLLOs in carbon-chain and organic molecular lines should shed light on
the chemical evolutional status of dense gas surrounding these VeLLOs and the initial stage
of star formation.

\section{Observations}

Observations of L1521F-IRS and IRAM 04191 were carried out
in the carbon-chain and organic molecular lines listed in Table 1
with the 45 m telescope at Nobeyama Radio Observatory\footnote{Nobeyama
Radio Observatory (NRO) is a branch of
the National Astronomical Observatory, an inter-university research
institute operated by the Ministry of Education, Culture, Sports,
Science and Technology of Japan.} from 10th to 18th on March, 2008.
All the molecular lines listed in Table 1 were observed simultaneously with two
SIS-mixer receivers (S80 and S100), S80 for the 85 GHz lines and S100 for the
110 GHz lines. The beam squint between the two receivers
was confirmed to be less than 1$\arcsec$. Typical single-sideband
(SSB) system noise temperatures were $\sim$ 270 K and $\sim$ 330 K for S80 and S100,
respectively. The beam size and the main beam efficiency of the telescope were
18$\arcsec$ and 42$\%$ at 85 GHz and 15$\arcsec$ and 40$\%$ at 110 GHz.
The telescope pointing
was checked every 90 - 120 minutes by observing the SiO maser emission
from NML-Tau at 43 GHz, and was confirmed to be better than 5$\arcsec$.
The position-switching mode was adopted for our observations.
A bank of eight high-resolution acousto-optical spectrometers
(AOS) with a bandwidth of 40 MHz and a frequency resolution of 37 kHz was
used as a backend.
The frequency resolution corresponds to 0.13 \kms at 85 GHz and 0.10 \kms
at 110 GHz.

Our observations consisted of two parts; one-point observations towards L1521F-IRS
($\alpha$$_{2000}$ = 04$^{\rm h}$ 28$^{\rm m}$ 38$\fs$95,
$\delta$$_{2000}$ = 26$^{\circ}$ 51$\arcmin$ 35$\farcs$1) and IRAM 04191
($\alpha$$_{2000}$ = 04$^{\rm h}$ 21$^{\rm m}$ 56$\fs$9,
$\delta$$_{2000}$ = 15$^{\circ}$ 29$\arcmin$ 46$\farcs$4), and mapping observations
around these two sources. The one-point observations were conducted
every observing day to calibrate the intensity variation caused by the receiver
gains and/or sky conditions and to obtain high signal-to-noise ratio spectra
with a long integration time. The resultant total on-source integration times
over the whole observing day were 39.7 minutes for L1521F IRS and 43.7 minutes
for IRAM 04191. The intensity variation during the whole
observing period was confirmed to be within 20 $\%$. The results obtained from the
one-point observations are summarized in Figure 1 and Table 2.

The mapping observations were performed at a grid spacing of 10$\arcsec$
and position angles of 0$\degr$ and 30$\degr$ in L1521F and IRAM 04191,
respectively, to align the mapping grid to the axis of the associated outflow.
With a typical integration time of 3 minutes per pointing,
typical rms noise levels of $\sim$ 0.2 K at 85 GHz and $\sim$ 0.3 K
at 110 GHz in units of $T_{\rm MB}$ were obtained (hereafter we present line 
intensities in units of $T_{\rm MB}$).

\placetable{tbl-1}

\section{Results}
\subsection{Spectra}

\begin{figure}
\epsscale{1.0}
\plotone{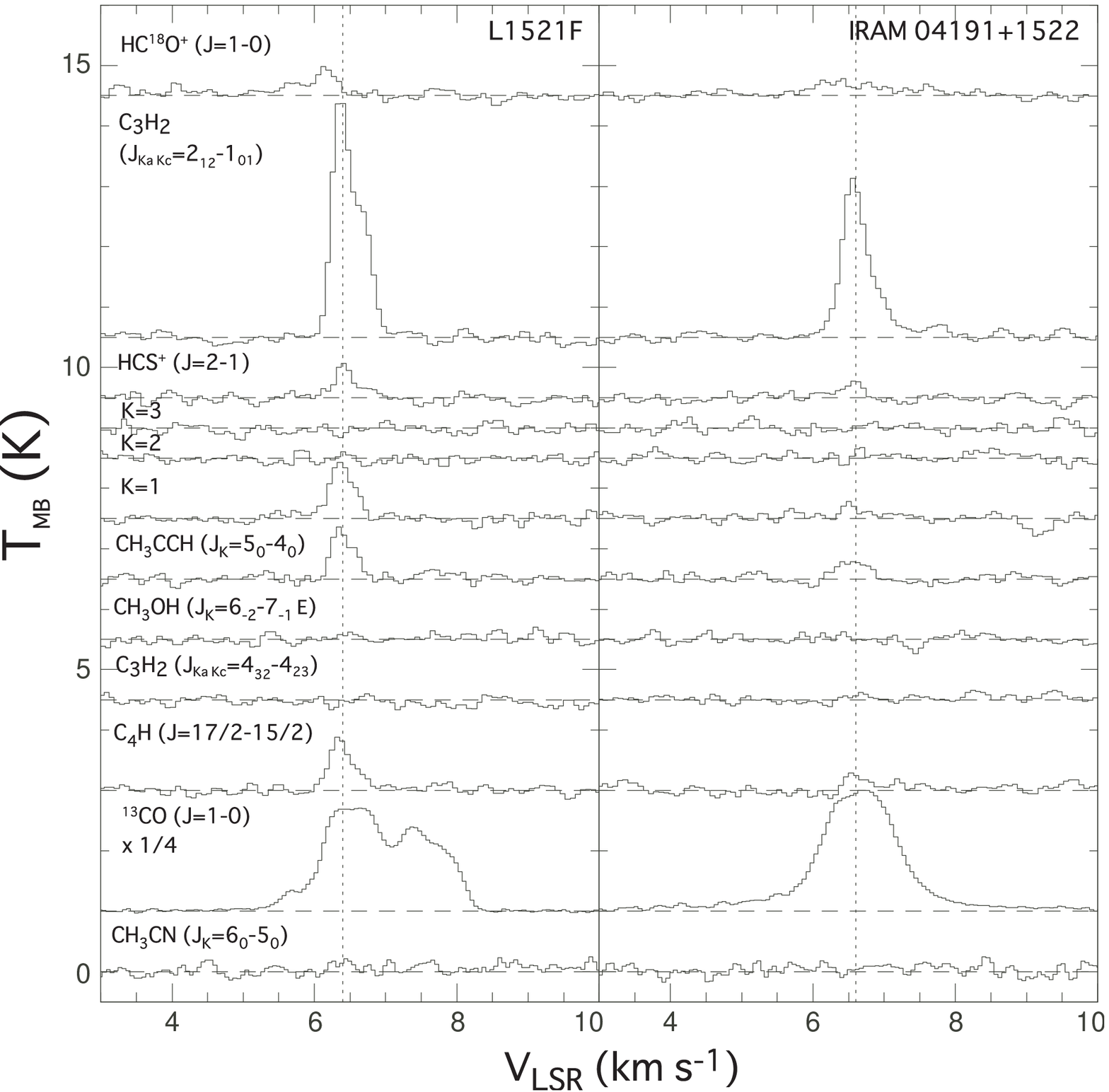}
\caption{Observed spectra towards L1521F-IRS ($Left$) and IRAM 04191 ($Right$).
Vertical dashed lines show the estimated systemic velocity of
6.4 km s$^{-1}$ and 6.6 km s$^{-1}$ for L1521F and IRAM 04191, respectively.
\label{spec}}
\end{figure}

Figure \ref{spec} shows profiles of the carbon-chain and organic molecular lines
towards L1521F-IRS and IRAM 04191, taken from the one-point observations.
All the molecular lines listed in Table 1,
except for the CH$_{3}$CN ($J_{K}$ = 6$_{K}$--5$_{K}$), and the higher-energy
CH$_{3}$OH ($J_K$ = 6$_{-2}$--7$_{-1}$ E), C$_{3}$H$_{2}$ ($J_{K_{a}K_{c}}$ = 4$_{32}$--4$_{23}$),
and the CH$_{3}$CCH ($J_{K}$ = 5$_{K}$--4$_{K}$; K $\geq$ 2) lines, were detected
above 3$\sigma$ in both sources.
Table 2 summarizes the observed line parameters,
derived from single-component Gaussian fittings to these spectra
unless otherwise noted.
It is clear that intensities of the carbon-chain molecular lines towards L1521F-IRS
are stronger than those towards IRAM 04191. The brightness temperatures
of the CH$_{3}$CCH ($J_{K}$ = 5$_{K}$--4$_{K}$; K = 0, 1) and
the C$_{4}$H ($N$ = 9--8, $J$ = $\frac{17}{2}$--$\frac{15}{2}$) lines towards L1521F-IRS
are $\sim$ 3 times more intense than those towards IRAM 04191. The C$_{3}$H$_{2}$
($J_{K_{a}K_{c}}$ = 2$_{12}$--1$_{01}$) line towards L1521F-IRS is also stronger
than that towards IRAM 04191.
The difference in the intensities of the carbon-chain molecular lines
is likely to reflect the different chemical compositions of the dense gas
around L1521F-IRS and IRAM 04191. We will discuss this point in more detail.


The $^{13}$CO (1--0) emission towards L1521F-IRS exhibits two velocity components
at $v_{LSR}$ = 6.6 km s$^{-1}$ and 7.4 km s$^{-1}$.
The 6.6 km s$^{-1}$ component with a stronger peak is detected at the same LSR velocity
where the other molecular lines are also detected, and most
probably traces the same cloud core component as that traced by the other
molecular lines. By contrast, the other molecular lines are not detected at the
LSR velocity where the 7.4 km s$^{-1}$ component is detected. The absence of the other
molecular lines is not due to chemistry because even the
C$^{18}$O (1--0) line does not show emission at this LSR velocity \citep{cra04},
suggesting that the 7.4 km s$^{-1}$ component probably traces a more diffuse cloud component.
L1521F is located in the western part of the L1521 molecular complex,
where the eastern edge of the redshifted (7 - 9 km s$^{-1}$) molecular filament
extending from the B213 and L1495 molecular complex overlaps \cite{gol08}.
The 7.4 km s$^{-1}$ component probably arises from this redshifted molecular filament.


\placefigure{spec}
\placetable{tbl-2}

\subsection{Spatial Distribution}
\subsubsection{L1521F}

\begin{figure}
\epsscale{0.8}
\plotone{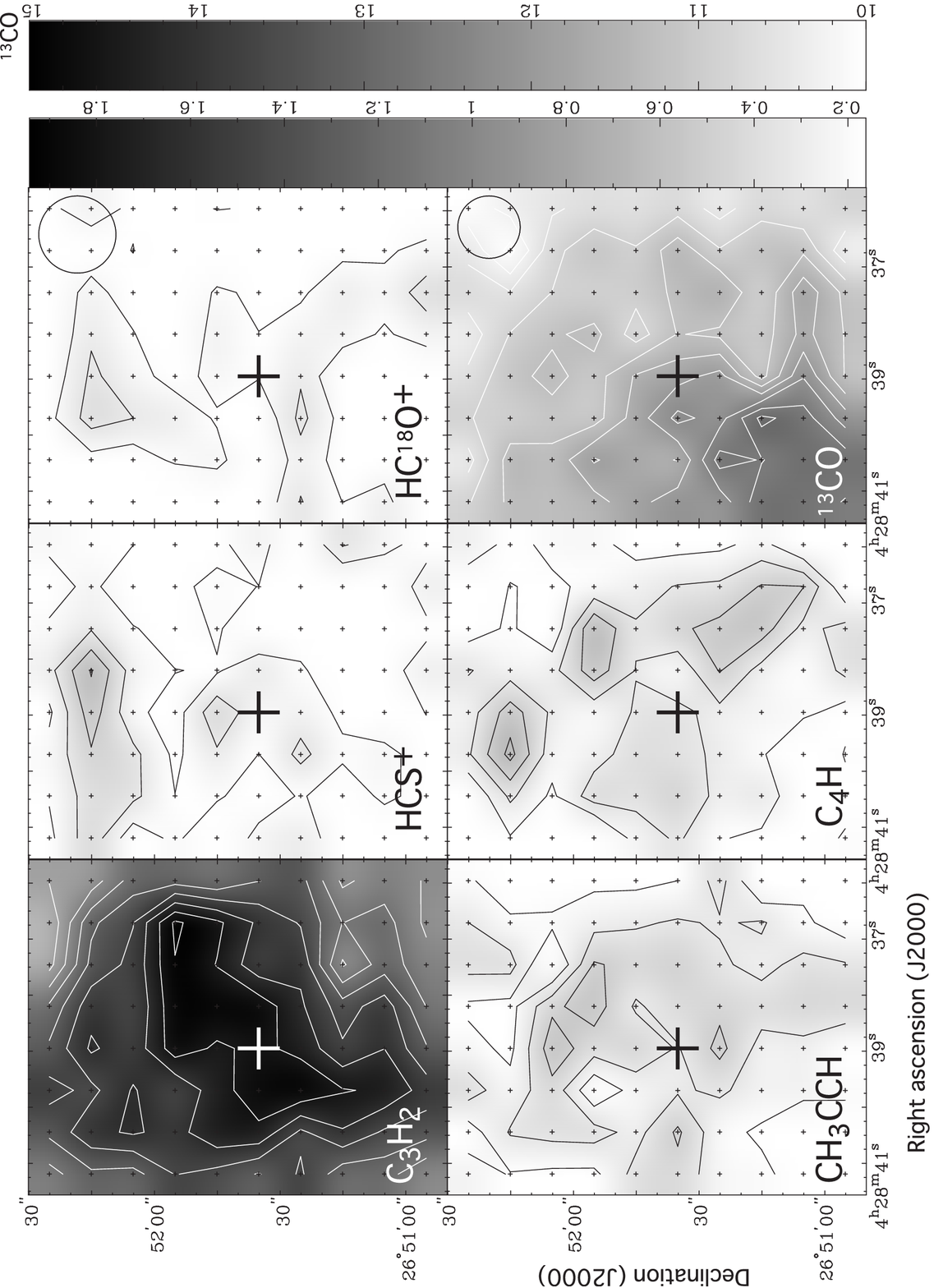}
\caption{Total integrated intensity maps of the C$_{3}$H$_{2}$
($J_{K_{a}K_{c}}$ = 2$_{12}$--1$_{01}$), CH$_{3}$CCH ($J_{K}$ = 5$_{0}$--4$_{0}$),
HCS$^{+}$ ($J$ = 2--1), C$_{4}$H ($N$ = 9--8, $J$ = $\frac{17}{2}$--$\frac{15}{2}$),
HC$^{18}$O$^{+}$ ($J$ = 1--0), and the $^{13}$CO ($J$ = 1--0) lines in L1521F.
The integrated velocity range is from 5.85 km s$^{-1}$ to 7.02 km s$^{-1}$ in the
C$_{3}$H$_{2}$, CH$_{3}$CCH, HCS$^{+}$, C$_{4}$H, and HC$^{18}$O$^{+}$ maps,
and from 4.9 km s$^{-1}$ to 8.4 km s$^{-1}$ in the $^{13}$CO map.
Crosses indicate the observed points, and large crosses show the protostellar position
(Spitzer Source; Bourke et al. 2006). Open circles in the HC$^{18}$O$^{+}$ and the $^{13}$CO panels
show the beam sizes for the 85 GHz and the 110 GHz observations, respectively.
Contour levels in the C$_{3}$H$_{2}$, CH$_{3}$CCH,  HCS$^{+}$, C$_{4}$H, and
HC$^{18}$O$^{+}$ maps are from 2$\sigma$ in steps of
2$\sigma$, where 1$\sigma$ is 0.081 K km s$^{-1}$.
The peak contour value in the C$_{3}$H$_{2}$ map at the north-west of the protostar is 24$\sigma$.
Contour levels in the $^{13}$CO map
are in steps of 2$\sigma$ (1$\sigma$ = 0.17 K km s$^{-1}$), where the peak contour value
at the south-east of the protostar is 76$\sigma$.
\label{l15mom0}}
\end{figure}

Figure \ref{l15mom0} shows total integrated intensity maps of the
C$_{3}$H$_{2}$ (2$_{12}$--1$_{01}$), HCS$^{+}$ (2--1), HC$^{18}$O$^{+}$ (1--0),
CH$_{3}$CCH (5$_{0}$--4$_{0}$), C$_{4}$H ($\frac{17}{2}$--$\frac{15}{2}$),
and the $^{13}$CO (1--0) lines in L1521F.
The map of the CH$_{3}$CCH (5$_1$ -- 4$_1$) line is not presented here because its distribution is
basically the same as that of the CH$_{3}$CCH (5$_0$ -- 4$_0$) line.
While the $^{13}$CO (1--0) emission line
shows an almost featureless distribution, other emission lines show more complicated
distributions. In this subsection, the spatial distributions of those
emission lines showing more complicated features are discussed first, and then the distribution
of the $^{13}$CO (1--0) emission line.

A common feature of the C$_{3}$H$_{2}$, HCS$^{+}$, HC$^{18}$O$^{+}$, CH$_{3}$CCH, and
the C$_{4}$H maps is that there is no emission peak at the position of L1521F-IRS.
In fact, the CH$_{3}$CCH (5$_{0}$--4$_{0}$), C$_{4}$H ($\frac{17}{2}$--$\frac{15}{2}$),
and the HCS$^{+}$ (2--1) lines show local emission minima close to the protostellar position.
On the other hand, detailed comparison among the different molecular lines shows that
their spatial distributions are different from each other.
For example, the C$_{3}$H$_{2}$ (2$_{12}$--1$_{01}$) line
shows the highest peak at (-30$\arcsec$, +20$\arcsec$), where there is no emission peak in the other molecular lines.
Although several emission peaks are seen in the CH$_{3}$CCH (5$_{0}$--4$_{0}$)
and C$_{4}$H ($\frac{17}{2}$--$\frac{15}{2}$) lines, their peak positions are
different from each other.
The distributions of the HCS$^{+}$ and HC$^{18}$O$^{+}$ lines
are also different from those of the carbon-chain lines.

\begin{figure}
\epsscale{0.55}
\plotone{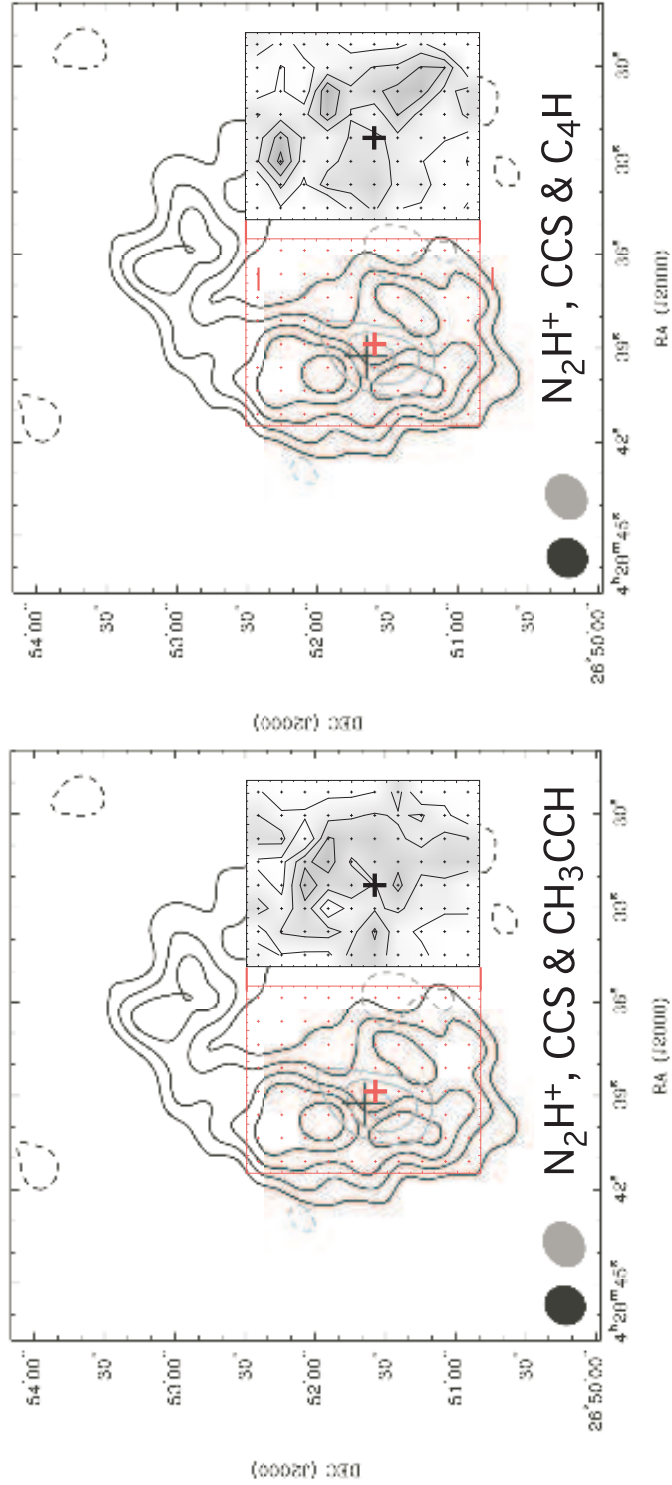}
\caption{Comparison of carbon-chain and N$_{2}$H$^{+}$ emission distributions in L1521F.
The distributions of the CCS (3$_2$--2$_1$) and N$_{2}$H$^{+}$ (1--0) emission are shown
in black and gray contours, respectively, taken from Figure 2 by Shinnaga et al. (2004).
Red rectangles, large and small crosses overlaid show our observing region,
the Spitzer source position and our observing points,
respectively. The maps of the CH$_{3}$CCH ($left$) and C$_{4}$H ($right$) emission distribution shown in
Figure \ref{l15mom0} are inserted.\label{hiroko}}
\end{figure}

In order to investigate the nature of the observed carbon-chain emission
distributions in more detail,
in Figure \ref{hiroko} we compare the CH$_{3}$CCH and C$_{4}$H emission
distributions to the CCS ($J_N$ = 3$_{2}$--2$_{1}$) and N$_{2}$H$^{+}$ ($J$ = 1--0)
distributions observed by Shinnaga et al. (2004).
The spatial resolution of the CCS and N$_{2}$H$^{+}$ observations
($\sim$ 18$\arcsec$) is similar to that of our observations, allowing us to
directly compare the spatial distribution of these molecular lines.
The N$_{2}$H$^{+}$ emission shows a compact ($\sim$ 50$\arcsec$ $\times$ 30$\arcsec$)
blob around the VeLLO position, while the CCS emission shows a local minimum
at the VeLLO position. The CH$_{3}$CCH and C$_{4}$H emissions
also appear to show local minima at the VeLLO position.
The exact peak positions of these carbon-chain emissions in L1521F, however,
are different from each other. Towards the position of the strongest northern CCS peak
the CH$_{3}$CCH and C$_{4}$H lines appear to show local emission minima, and there is neither
a CH$_{3}$CCH nor a C$_{4}$H counterpart to the south-eastern CCS peak.


Compared to these carbon-chain molecular emissions,
the $^{13}$CO (1--0) emission shows a rather featureless distribution, which
is also seen in the C$^{18}$O (1--0, 2--1) emission \cite{cra04}. 
From the C$^{18}$O (1--0) and C$^{17}$O (1--0) observations
Crapsi et al. (2004) estimated the optical depth of the
C$^{18}$O (1--0) line in L1521F to be $\sim$ 1.5.
On the assumption of the abundance
ratio of [$^{13}$CO]/[C$^{18}$O] of 7.7 \cite{wil94}
the optical depth of the
$^{13}$CO (1--0) emission is $\sim$ 12.
The featureless emission distribution in the $^{13}$CO (1--0)
line is probably due to this high optical depth.
Although near-infrared observations of L1521F with
Spitzer found a bipolar nebulosity oriented along the east-west
direction (P.A. $\sim$ 90$\degr$) around the protostellar source,
implying the presence of an outflow \cite{bou06},
no high-velocity ($>$ 2 km s$^{-1}$) winglike component is seen
in the $^{13}$CO (1--0) velocity channel maps at an rms noise level of $\sim$ 0.29 K.

\placefigure{l15mom0}
\placefigure{hiroko}

\subsubsection{IRAM 04191}

\begin{figure}
\epsscale{0.7}
\plotone{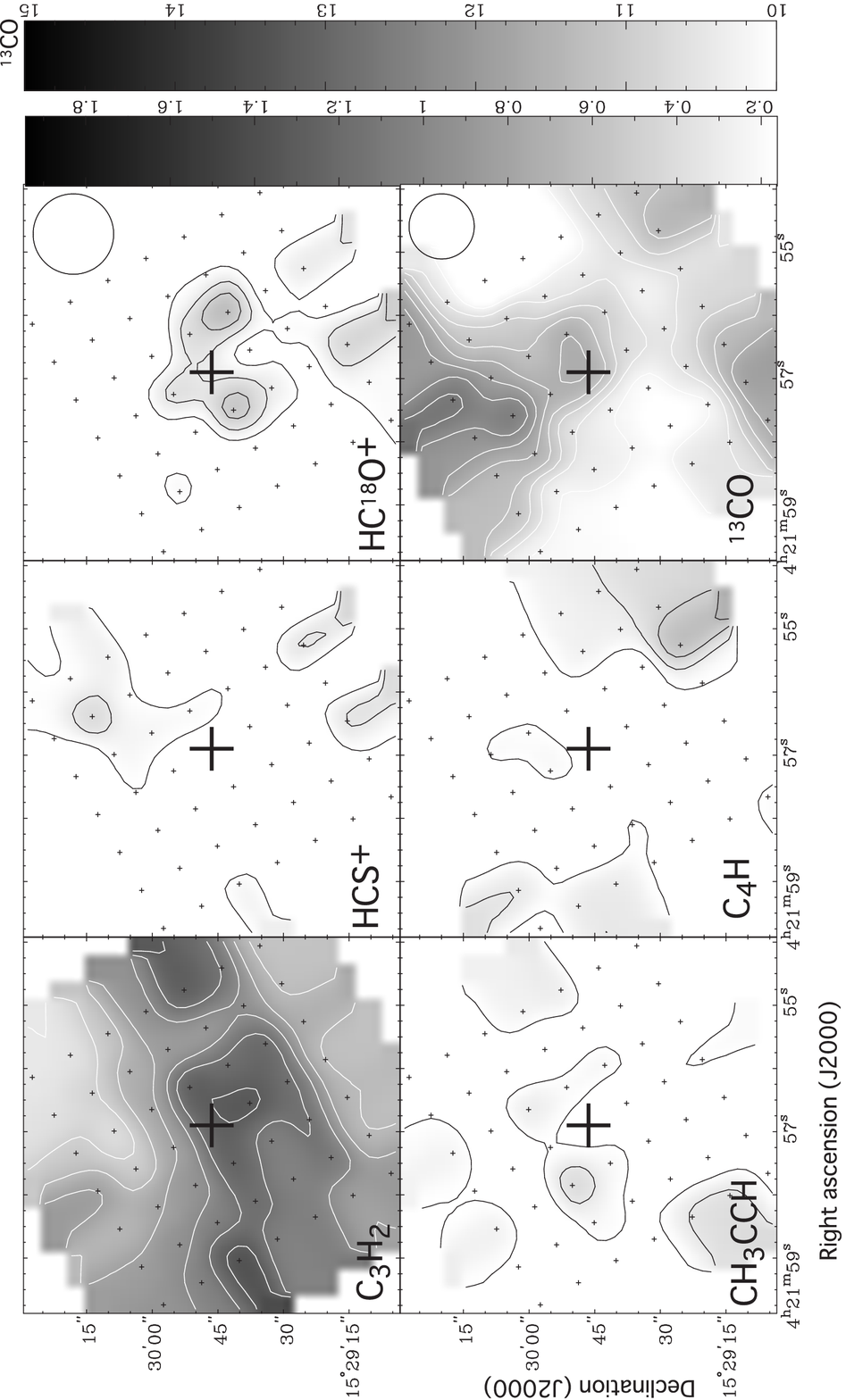}
\caption{Total integrated intensity maps of the C$_{3}$H$_{2}$
($J_{K_{a}K_{c}}$ = 2$_{12}$--1$_{01}$),
CH$_{3}$CCH ($J_{K}$ = 5$_{0}$--4$_{0}$), HCS$^{+}$ ($J$ = 2--1),
C$_{4}$H ($N$ = 9--8, $J$ = $\frac{17}{2}$--$\frac{15}{2}$),
HC$^{18}$O$^{+}$ ($J$ = 1--0), and the $^{13}$CO ($J$ = 1--0) lines in IRAM 04191.
The integrated velocity range is from 5.46 km s$^{-1}$ to 7.54 km s$^{-1}$ in the
C$_{3}$H$_{2}$, CH$_{3}$CCH, HCS$^{+}$, C$_{4}$H, and the HC$^{18}$O$^{+}$ maps,
and from 4.1 km s$^{-1}$ to 10.1 km s$^{-1}$ in the $^{13}$CO map.
Crosses indicate the observed points, and large crosses show the protostellar position
\cite{an99a}. Open circles in the HC$^{18}$O$^{+}$ and the $^{13}$CO panels
show the beam sizes for the 85 GHz and the 110 GHz observations, respectively.
Contour levels are the same as in Figure \ref{l15mom0}.
The peak contour value in the C$_{3}$H$_{2}$ map near the protostellar position
is 16$\sigma$, and that in the $^{13}$CO map at the northern edge is 74$\sigma$.
\label{i04mom0}}
\end{figure}

Figure \ref{i04mom0} shows the same total integrated intensity maps
as Figure \ref{l15mom0} but for IRAM 04191. Compared to L1521F,
the intensities of the C$_{3}$H$_{2}$ (2$_{12}$--1$_{01}$),
CH$_{3}$CCH (5$_{0}$--4$_{0}$), C$_{4}$H ($\frac{17}{2}$--$\frac{15}{2}$),
and the HCS$^{+}$ (2--1) lines are significantly weaker in IRAM 04191.
Among these lines only the C$_{3}$H$_{2}$ line is strong enough
to investigate the spatial distribution.
We note that both L1521F-IRS and IRAM 04191
are associated with intense
and extended ($>$ 4000 AU) dust-continuum emission,
and that the amount of molecular gas
derived from the dust-continuum emissions around these sources is
similar \cite{an99a,an99b,mot01}.
Hence, these weak carbon-chain molecular emissions in IRAM 04191 are mostly likely due to the
lower abundances of these molecules as will be discussed in $\S$3.3.
On the other hand, the peak total integrated intensity of the HC$^{18}$O$^{+}$ (1--0)
emission in IRAM 04191 is more than twice as high as that in L1521F.
Two HC$^{18}$O$^{+}$ emission peaks are seen to the south-east and
south-west of the protostar with a local minimum at the protostellar position.

The distribution of the C$_{3}$H$_{2}$ emission in IRAM 04191 observed
with the 45 m telescope is consistent with the earlier result
observed with the IRAM 30 m telescope \cite{bel02}.
The emission shows a peak close to the protostar,
and an elongated feature along the north-west to the
south-east direction, which is approximately perpendicular to the direction
of the associated outflow \cite{an99a,lee02}.
This elongated structure most likely traces the circumstellar envelope around
the protostar, also seen in other molecular lines such as N$_{2}$H$^{+}$ (1--0)
\cite{bel02}, C$^{34}$S (2--1) and CH$_{3}$OH (2--1) \cite{tak03}.
Compared to the C$_{3}$H$_{2}$ emission the $^{13}$CO (1--0) emission
shows quite a different distribution; the $^{13}$CO emission shows fan-shaped structures
at the south-west and north-east of the protostar, and
the south-western and north-eastern $^{13}$CO (1--0) emissions
are blueshifted ($V_{LSR}$ = 4.1 - 5.7 \kms) and redshifted (7.7 - 10.1 \kms), respectively.
Since the fan-shaped structures and the blueshifted and redshifted sense of the
$^{13}$CO (1--0) emission are similar to those of the
associate molecular outflow observed in the $^{12}$CO emissions \cite{lee02,lee05},
the $^{13}$CO (1--0) emission is thought to primarily trace
the associated molecular outflow.


\placefigure{i04mom0}

\subsection{Molecular Column Densities and Abundances}

On the assumption of optically-thin emission and
LTE conditions, the column densities of the observed molecules ($N_{mol}$)
can be estimated as follows:

\begin{equation}
N_{mol}=\frac{8\pi\nu^{3}}{c^{3}}\frac{1}{g_{u}A}
\frac{Z({T_{\rm ex}})}
{\exp(-\frac{E_u}{kT_{\rm ex}})(\exp(\frac{h\nu}{kT_{\rm ex}})-1)}
\frac{\int T_{MB} dv}{J(T_{\rm ex})-J(T_{\rm bg})},
\end{equation}
where
\begin{equation}
J(T)=\frac{\frac{h\nu}{k}}{\exp(\frac{h\nu}{kT})-1}. 
\end{equation}

In the above expressions, $h$ is the Planck constant, $k$ 
is Boltzmann's constant, $c$ is the speed of light, $\nu$ is the line 
frequency, $T_{\rm ex}$ is the excitation temperature, $T_{\rm bg}$ is
the background radiation temperature, $A$ is the Einstein $A$-coefficient,
$Z({T_{\rm ex}})$ is the partition function, $E_u$ is the rotational
energy level of the upper energy state, $g_u$ is the statistical weight
of the upper energy state, and $\int T_{MB} dv$
is the integrated line intensity. Values of $\nu$, $A$, $E_u$ and $g_u$
are listed in Table 1. Values of the partition functions were taken from
the JPL molecular catalog
on the assumption of $T_{\rm ex}$ = 9.375 K  \cite{pic98}
for all the molecular lines both in L1521F and IRAM 04191,
since C$^{18}$O (1--0) and C$^{17}$O (1--0) observations in L1521F \cite{cra04}
and CH$_{3}$OH (2--1, 5--4) observations in IRAM 04191 \cite{tak03}
have shown that the excitation temperatures are $\sim$ 10 K.
With an excitation temperature of 9.375 K the optical depth of the
C$_{3}$H$_{2}$ (2$_{12}$--1$_{01}$) emission towards L1521F-IRS,
the most intense emission except for the $^{13}$CO (1--0) emission,
is estimated to be $\sim$ 0.95, and all of the other, weaker, molecular emissions
are optically thin. It is possible that the excitation temperatures of
some of the observed molecular lines are lower than 9.375 K,
and that some of the observed molecular lines such as the
C$_{3}$H$_{2}$ (2$_{12}$--1$_{01}$) line are optically thick.
In such a case, the estimates of the molecular column densities should be
regarded as lower limits.
As already discussed in $\S3.2$ the $^{13}$CO (1--0)
emission is likely to be optically thick, and hence the
$^{13}$CO column density cannot be estimated.
Estimated molecular column
densities towards the protostellar position of L1521F-IRS and IRAM 04191,
derived from the $\int T_{B} dv$ values listed in Table 2, are shown in Table 3.

In order to estimate abundances of the observed molecules,
values of the column density of molecular hydrogen ($N_{\rm H_{2}}$) towards L1521F-IRS and IRAM 04191
are required.
For the estimate of $N_{\rm H_{2}}$ we adopted 1.2 mm dust-continuum emission
observed with the IRAM 30~m telescope towards L1521F-IRS \cite{cra04} and IRAM 04191 \cite{an99b,mot01},
and calculated the $N_{\rm H_{2}}$ values using the same formula
and the dust temperature (= 10 K) adopted by Crapsi et al. (2004).
The $N_{\rm H_{2}}$ values were estimated to be 9.26 $\times$ 10$^{22}$ cm$^{-2}$
and 1.27 $\times$ 10$^{23}$ cm$^{-2}$ towards
L1521F-IRS and IRAM 04191, respectively.
The molecular abundances ($X_{mol}$) can then be estimated as
$X_{mol}$ = $N_{mol}$ / $N_{\rm H_{2}}$. Here, the difference in the beam sizes between our observations
(18$\arcsec$) and the 1.3 mm continuum observations with the IRAM 30 m telescope
(13$\arcsec$) was not taken into account. The $X_{mol}$ values towards L1521F-IRS and IRAM 04191
are summarized in Table 3.
There is a clear difference in the molecular abundances between L1521F
and IRAM 04191. The C$_{4}$H and CH$_{3}$CCH abundances in L1521F are a factor of 5 and 3
higher than those in IRAM 04191, respectively. 
The C$_{3}$H$_{2}$ abundance in L1521F is also a factor of 2 higher
that in IRAM 04191.
These results show that carbon-chain molecules are more abundant in L1521F than IRAM 04191.
The HCS$^{+}$ abundance in L1521F is also a factor of 3 higher that
in IRAM 04191. On the other hand, the HC$^{18}$O$^{+}$ abundance in L1521F is similar to
or possibly lower than that in IRAM 04191.

\placetable{tbl-3}

\section{Discussion}
\subsection{Chemical Evolutionary Difference between L1521F and IRAM 04191}

As shown in the previous section, our observations of the carbon-chain molecules show clear
chemical differentiation between L1521F and IRAM 04191.
The chemical differentiation could reflect the evolutionary difference between the embedded VeLLOs
of L1521F-IRS and IRAM 04191.
Here, we will discuss the link between the chemical differences and the evolutionary difference
in these VeLLOs.

Chemistry in dense molecular-cloud cores is known to be non-equilibrium and
time-dependent, and molecular abundances in dense cores change as a function of the evolutionary
time \cite{leu84,her89,ber97}.
In the gas phase, the time dependences of molecular abundances
are mainly controlled by the status of carbon.
At an early evolutionary stage ($\leq$ 10$^{5}$ yr) carbon is mainly
in the form of C$^{+}$ and C,
and at late evolutionary stages ($>$ 10$^{5}$ yr) almost all carbon
is in the form of CO.
Carbon-chain molecules are produced through hydrogenation of C$^{+}$ and C,
and hence are more abundant in the earlier evolutionary stage.
On the other hand, HCO$^{+}$ is formed via CO + H$_{3}^{+}$,
and HC$^{18}$O$^{+}$ is more abundant at later evolutionary phases.
Thus, the higher abundances of the carbon-chain molecules in L1521F than in IRAM 04191,
as well as the sightly higher HC$^{18}$O$^{+}$ abundance in IRAM 04191 than in L1521F,
are likely to indicate that
IRAM 04191 is chemically more evolved than L1521F.
This chemical evolutionary difference
is also consistent with the different outflow activity between L1521F-IRS and IRAM 04191;
there is a well-developed CO molecular outflow in IRAM 04191, as seen in Figure \ref{i04mom0},
while no clear CO molecular outflow is observed in L1521F-IRS,
suggesting that the star-forming activity is more advanced in IRAM 04191
than L1521F-IRS.

Aikawa et al. (2001, 2003, 2005) theoretically studied the evolution of
molecular distributions in collapsing dense cores.
Their model shows that the
radial distribution of molecules varies as a function of the evolutionary
time, and that the distribution of carbon-chain molecules such as CCS first shows
a central hole after $\sim$ 10$^{6}$ yr (with $\alpha$ = 1.1,
where $\alpha$ denotes the internal gravity-to-pressure ratio in the collapsing
dense core.). This is because
at the center the gas density is higher than that at the outer part,
and hence the chemical evolution proceeds faster than that at the outer part.
On the other hand ``late-type'' molecules such as CO, HCO$^{+}$, and N$_{2}$H$^{+}$
show centrally-peaked column density profiles in the early evolutionary stage, and
then CO and HCO$^{+}$ start to show
central holes due to the depletion onto grain surfaces
($>$ 10$^{6}$ yr).
Eventually, N$_{2}$H$^{+}$ also becomes depleted \cite{ber02,bel04}.
In L1521F, the CH$_{3}$CCH, C$_{4}$H, and the CCS emissions  do not show a centrally-peaked spatial
distribution but the N$_{2}$H$^{+}$ emission shows a centrally-peaked distribution
(see Figure \ref{hiroko}). On the other hand, in IRAM 04191
the CH$_{3}$CCH and C$_{4}$H emissions are as weak as our detection limit,
and the depletion of N$_{2}$H$^{+}$ in the central $\sim$ 1600 AU has been reported \cite{bel04}.
Comparison between the theoretical model and these observational results
implies that in L1521F carbon-chain molecules are still
abundant but at the center,
they become less abundant due to the faster gas-phase chemical evolution, while
in IRAM 04191 these carbon-chain molecules become undetectable
and the depletion of N$_{2}$H$^{+}$ proceeds at the center.



\subsection{Comparison with Other Dense Cores}

In $\S$4.1., we have discussed the possibility that L1521F-IRS is in an earlier evolutionary stage than IRAM 04191.
It is interesting to then
compare the evolutionary stage of the surrounding dense gas around L1521F-IRS and IRAM 04191
to that of starless dense cores in Taurus.
In Table 4, we compiled several evolutionary indicators
of three starless cores (L1521B, L1498, and L1544) and dense cores around the two VeLLOs
(L1521F and IRAM 04191). The evolutionary indicators include
the ratio between the N$_{2}$H$^{+}$ and CCS column densities, the deuterium
fractionation measured from the N$_{2}$D$^{+}$ / N$_{2}$H$^{+}$
column density ratio, the CO depletion factor defined by
Crapsi et al. (2004), the central molecular-gas
density, and the CH$_{3}$CCH and C$_{4}$H column densities.
As already discussed in $\S$4.1., the ratio between the 
N$_{2}$H$^{+}$ and CCS column densities is an excellent
indicator of the gas-phase chemical evolution in dense cores, 
and the ratio increases as the dense core evolves.
The deuterium fractionation and the CO depletion factor
are measures of chemical evolution of cold dense gas.
Under the cold gas ($\leq$ 10 K) condition the deuterium fractionation
proceeds due to preferable condition for the reaction
of H$_{3}^{+}$ + HD $\rightarrow$ H$_{2}$D$^{+}$ + H$_{2}$,
and CO molecules keep
depleting onto grain surfaces \cite{cra04,cr05a}.
The central gas density is a
direct measure of the physical evolutionary stage of dense cores.

Table 4 shows that the collected dense cores can be sorted following these
evolutional criteria.
In L1521B and L1498 the N$_{2}$H$^{+}$ to CCS ratio, the deuterium fractionation,
the CO depletion factor, and the central
gas density are lower than those in L1521F and IRAM 04191.
The CO depletion factor and the central gas density in L1521F
are similar to those in L1544, another starless core in Taurus \cite{oha99,cr05a},
but the deuterium fractionation is twice as high as in L1544 than in L1521F.
These results imply that L1544 is in a similar evolutionary stage as
L1521F. In fact, towards L1544 an infalling gas motion has been observed \cite{wil99,oha99},
suggesting the core is in the process of protostellar formation, though inspection of the
Spitzer data in L1544 does not reveal any protostellar candidate \cite{bou06}.
IRAM 04191 shows a higher N$_{2}$H$^{+}$ to CCS ratio, and CO depletion factor,
and lower carbon-chain abundances than L1521F, and
IRAM 04191 is probably more evolved than L1521F as already discussed.

\placetable{tbl-4}

\subsection{Molecular Desorption}

Once a protostar forms at the center of the dense core,
heating from the protostar desorbs molecules from
grain surfaces and alters the chemical conditions.
Aikawa et al. (2008) showed that the warm molecular region heated by the protostar
expands during protostellar evolution, and that molecules
such as CO are desorbed from grain surfaces and show
abundance enhancement within the radius where the gas
temperature exceeds their sublimation temperatures.
Organic molecules such as CH$_{3}$OH and CH$_{3}$CN
formed on grain surfaces are also desorbed and enhanced in the gas phase
by orders of magnitude within the radius where the gas
temperature is above $\sim$ 100 K ($\equiv$ $R_{100}$).

The model by Aikawa et al. (2008) predicts that
$R_{100}$ reaches 100 (AU) after $\sim$ 10$^{5}$ years
from protostellar formation, and that
the mean CH$_{3}$OH abundance ($\equiv$ $X_{\rm CH_3OH}$)
and the mean gas kinetic temperature ($\equiv$ $T_{\rm K}$)
averaged over the 45-m beam ($\sim$ 18$\arcsec$ $\sim$ 2500 AU) are $\sim$ 8.0 $\times$ 10$^{-9}$
and $\sim$ 52 K, respectively.
On the other hand, N$_{2}$H$^{+}$ and 1.3 mm dust-continuum observations
of L1521F \cite{cra04} and IRAM 04191 \cite{bel04} show that
the gas density towards the central
$\sim$ 20$\arcsec$ region ($\equiv$ $n_{\rm H_2}$) is $\sim$ 10$^{6}$ cm$^{-3}$
and that the velocity width over the core size ($\equiv$ $dv/dr$) is
$\sim$ 10 km s$^{-1}$ pc$^{-1}$.
With input values of $n_{\rm H_2}$ = 10$^{6}$ cm$^{-3}$, $T_{\rm K}$ = 52 K,
and $X_{\rm CH_3OH} / dv/dr$ = 8.0 $\times$ 10$^{-10}$ km$^{-1}$ s pc,
our LVG calculation of CH$_{3}$OH \cite{tak98,tak00} predicts the expected
CH$_{3}$OH (6$_{-2}$--7$_{-1}$ E) line intensity to be $\sim$ 0.63 K, which is significantly
above the 3$\sigma$ upper limit ($\sim$ 0.2 K) of our observations.
This LVG calculation suggests that $R_{100}$ in L1521F and IRAM 04191
has not yet reached 100 AU, and that
the evolutionary stage of L1521F-IRS and IRAM 04191 is younger than 10$^{5}$ years
from protostellar formation.
%
Similarly, Aikawa et al. (2008)
showed that at $R_{100}$ = 100 (AU) the beam-averaged CH$_{3}$CN abundance
reaches on the order of 10$^{-10}$, while the observed upper limit of
the CH$_{3}$CN abundance is $\sim$ 4 $\times$ 10$^{-13}$.
We consider that
the non-detection of these organic molecular lines is a sign
of the youthfulness of these VeLLOs.

Although there is no other observation in the 6$_{-2}$--7$_{-1}$ E
transition of CH$_{3}$OH towards low-mass protostars reported,
interferometric observations of Class 0 protostars have found
compact ($\leq$ 500 AU) millimeter and submillimeter CH$_{3}$OH emission
associated with the central protostars in L1157 \cite{gol99,vel02},
NGC1333 IRAS 2A \cite{jo05b}, and in IRAS 16293-2422 \cite{kua04,cha05}.
Similar compact components are also seen
in the CH$_{3}$CN (6--5) \cite{bot04} and CH$_{3}$CN (12--11) emission \cite{bis08}
towards IRAS 16293-2422.
The CH$_{3}$OH abundances in those compact components were estimated to be
$\sim$ 10$^{-8}$ towards L1157 \cite{gol99,vel02}, $\sim$ 3 $\times$ 10$^{-8}$
towards NGC1333 IRAS 2A \cite{jo05b}, and $\sim$ 9.4 $\times$ 10$^{-8}$
towards IRAS 16293-2422 \cite{cha05},
which are one order of magnitude higher than the CH$_{3}$OH abundance
in cold dark clouds ($\sim$ 2 $\times$ 10$^{-9}$) \cite{fri88,tak98,tak00}.
Maret et al. (2005) have conducted a CH$_{3}$OH (5$_{K}$--4$_{K}$; 7$_{K}$--6$_{K}$)
survey of seven Class 0 sources with the IRAM 30 m and JCMT telescopes.
From their radiative transfer and ``jump'' models,
they have found that four sources (IRAS 16293-2422, NGC1333 IRAS 2, IRAS4B, and L1448-MM)
show CH$_{3}$OH abundance enhancements up to 1-7 $\times$ 10$^{-7}$ at the
innermost part of the envelopes.
Similar single-dish studies of the CH$_{3}$CN lines have revealed
more than 2 orders of magnitude of abundance enhancements of CH$_{3}$CN
in NGC 1333 IRAS 2 (7 $\times$ 10$^{-9}$) \cite{jo05a}
and IRAS 16293-2422 (10$^{-8}$) \cite{caz03}.
These results show that towards several Class 0 protostars
the CH$_{3}$OH and CH$_{3}$CN abundance enhancements take place.

On the other hand, non-detection of the CH$_{3}$OH (6$_{-2}$--7$_{-1}$ E) and
CH$_{3}$CN (6$_{K}$--5$_{K}$) lines towards L1521F-IRS and IRAM 04191
does not imply the absence of the desorption region.
If $R_{100}$ = 10 AU for example,
$R_{20}$, the radius of the CO desorption, reaches 250 AU \cite{mas98,mas00}.
Our recent observations of the C$^{18}$O (2--1) line towards L1521F-IRS
and IRAM 04191 with the SMA have revealed compact ($\sim$ 500 AU) CO emission
associated with these VeLLOs, suggesting the possible desorption of CO.
We suggest that the VeLLOs of L1521F-IRS and IRAM 04191 are in the ongoing stage
of the desorption processes, and that the abundance enhancement of
organic molecules of CH$_{3}$OH and CH$_{3}$CN can be used
as a chemical evolutionary indicator from the VeLLO to Class 0 stage.

%


\subsection{Different Origin of L1521F-IRS and IRAM 04191 ?}

In the above discussion we implicitly assume that both L1521F-IRS and IRAM 04191
are protostars in a common evolutionary sequence and will form a star with a similar final mass.
On the other hand, our recent SMA observations of L1521F-IRS have revealed
compact ($\sim$ 1500 AU) molecular outflows \citep{tak10}, and the outflow properties are similar to those
around very low-mass stars or brown dwarfs \citep{nat04,pha08}.
It is therefore possible that L1521F-IRS is a very low-mass, more evolved star
than protostars. In this final subsection, we will discuss this alternative possibility.

The estimated outflow mass around L1521F-IRS is $\sim$ 3.6 $\times$ 10$^{-5}$ $M_{\odot}$,
which is orders of magnitude lower than that of IRAM 04191 ($\sim$ 3 $\times$ 10$^{-2}$ $M_{\odot}$;
Lee et al. 2002).
There are several possible explanations for the difference in the outflow masses.
One is that both L1521F-IRS and IRAM 04191 are protostars and 
the mass ejection event of L1521F-IRS has started more recently than IRAM 04191,
and hence the amount of the molecular material entrained by L1521F-IRS is lower than that
by IRAM 04191. This interpretation is consistent with the arguments in $\S$4.1 - $\S$4.3.
The second interpretation is that the mass accretion rate towards L1521F-IRS is intrinsically smaller
than that towards IRAM 04191, presumably due to the different effective sound speed between
the L1521F and IRAM 04191 region. The lower mass accretion rate towards L1521F-IRS could
yield the lower total outflow mass and the lower internal luminosity as compared to those of IRAM 04191.
The other possible interpretation is that L1521F-IRS is a very low-mass, more evolved star close to
the end of the mass accretion and ejection phase, and that the amount of the total mass of the
molecular outflow is proportional to the amount of the total accreted material, and hence
the central stellar mass.
In this case, there may be a substantial difference in the central stellar mass between L1521F-IRS and IRAM 04191,
and the most of the luminosity of L1521F-IRS may originate from nuclear burning whereas the luminosity
of IRAM 04191 originate mostly from mass accretion \cite{nat04}.

In this case, how can we explain the observed chemical difference of the surrounding
dense gas between L1521F and IRAM 04191 ?
As discussed in $\S$3.1, L1521F is located in the main ridge
of the Taurus Molecular Cloud complex \cite{dam01,gol08},
and there is ample molecular material surrounding the L1521F region.
On the other hand, IRAM 04191 is not located in the main Taurus Complex,
but at the south-western edge of the Taurus-Auriga Complex away from the
Galactic Plane ($b$ $\sim$ -23.5$\degr$) \cite{dam01}.
The surrounding ambient gas can be a source of the ``chemically-fresh'' molecular gas
and maintain the higher carbon-chain abundances around L1521F,
whereas around IRAM 04191 there is less such
molecular gas. The observed chemical difference
between L1521F and IRAM 04191 may be due to the different surrounding environment.

\section{Summary}

We have carried out mapping observations around the VeLLOs L1521F-IRS and
IRAM 04191+1522 in C$_{3}$H$_{2}$ ($J_{K_{a}K_{c}}$ = 2$_{12}$--1$_{01}$;
4$_{32}$--4$_{23}$),
CH$_{3}$CCH ($J_{K}$ = 5$_{K}$--4$_{K}$),
C$_{4}$H ($N$ = 9--8, $J$ = $\frac{17}{2}$--$\frac{15}{2}$),
HCS$^{+}$ ($J$ = 2--1), HC$^{18}$O$^{+}$ ($J$ = 1--0),
CH$_{3}$OH ($J_{K}$ = 6$_{-2}$--7$_{-1}$ E), CH$_{3}$CN ($J_{K}$ = 6$_{K}$--5$_{K}$),
and $^{13}$CO ($J$ = 1--0) lines with the Nobeyama 45-m telescope.
Our high-sensitivity observations
of these carbon-chain and organic molecular lines have provided
the following new results:

1. We detected all but the C$_{3}$H$_{2}$ (4$_{32}$--4$_{23}$),
CH$_{3}$OH (6$_{-2}$--7$_{-1}$ E), and the CH$_{3}$CN (6$_{K}$--5$_{K}$) lines
towards both L1521F-IRS and IRAM 04191 above the 3$\sigma$ upper limit of $\sim$ 0.2 K.
The intensities of the detected carbon-chain molecular lines,
C$_{3}$H$_{2}$ (2$_{12}$--1$_{01}$), CH$_{3}$CCH (5$_{K}$--4$_{K}$; K=0, 1),
and C$_{4}$H ($\frac{17}{2}$--$\frac{15}{2}$), are 1.5 to 3.5 times stronger
towards L1521F-IRS than IRAM 04191+1522.
The abundances of these carbon-chain molecules towards L1521F-IRS are 2 to 5 times as high
as those towards IRAM 04191+1522, while the HC$^{18}$O$^{+}$ abundance
towards L1521F-IRS is $\sim$ a factor of 1.5 lower than that towards IRAM 04191+1522.
These results suggest a different chemical status in the dense gas around L1521F-IRS and
IRAM 04191+1522.

2. Our mapping observations of L1521F have found that
the carbon-chain molecular emissions of C$_{3}$H$_{2}$ (2$_{12}$--1$_{01}$),
CH$_{3}$CCH (5$_{K}$--4$_{K}$) and
C$_{4}$H ($\frac{17}{2}$--$\frac{15}{2}$), and
the HCS$^{+}$ (2--1) emission do not show emission peaks
at the protostellar position
as seen in the CCS (3$_{2}$--2$_{1}$) emission.
Furthermore, each molecular line shows a different morphology
with different emission peaks.
No extended high-velocity ($>$ 2 km s$^{-1}$) $^{13}$CO (1--0) component
is seen in L1521F, suggesting that the molecular
outflow associated with L1521F-IRS is not yet well-developed.

3. Our mapping observations of IRAM 04191+1522 have revealed that,
in contrast to the results in L1521F,
the carbon-chain molecular lines are weak and that only
the C$_{3}$H$_{2}$ (2$_{12}$--1$_{01}$) line is intense enough
to trace the structure of the protostellar envelope.
The $^{13}$CO (1--0) emission shows
fan-shaped structure at the south-west and the north-east of IRAM 04191+1522,
and traces the molecular outflow driven by IRAM 04191+1522.

4. The systematic chemical differentiation found in the dense gas
around L1521F-IRS and IRAM 04191+1522
can be interpreted as different
evolutionary stages between these two VeLLOs; namely, IRAM 04191+1522
is more evolved than L1521F-IRS.
In L1521F, carbon-chain molecules, which are thought to be
abundant at the early evolutionary stage, are still
abundant but at the center
they are less abundant due to the faster gas-phase chemical evolution.
On the other hand in IRAM 04191+1522
these carbon-chain molecules have already been diminished.
The different evolutionary stages inferred from the chemical differentiation
is consistent with the extent of the associated molecular outflows
found in the $^{13}$CO (1--0) emission.
Comparison of the chemical evolutionary tracers of the carbon-chain and N$_{2}$H$^{+}$ column
densities, deuterium fractionation, CO depletion factor, and the central molecular-gas
density to those of starless cores in Taurus exhibits a systematic evolutionary trend.
It is also possible that the ample molecular material of the Taurus Molecular Cloud complex
keeps supplying ``chemically-fresh'' molecular gas to the L1521F core, while
the lack of such ambient molecular gas around IRAM 04191 prevents supply of such molecular
gas to the IRAM 04191 core.


5. The non-detection of the organic molecular lines of
CH$_{3}$OH (6$_{-2}$--7$_{-1}$ E) and CH$_{3}$CN (6$_{K}$--5$_{K}$)
implies that the radius of the warm ($\sim$ 100 K),
molecular-desorbing region heated by the central protostar is smaller than
$\sim$ 100 AU around L1521F-IRS and IRAM 04191+1522, suggesting a younger age
($<$ 10$^{5}$ yr) for these VeLLOs than that of the Class 0 sources associated
with the compact ($\lesssim$ 500 AU) CH$_{3}$OH emission.

\acknowledgments
We are grateful to S. Takahashi, J. Karr, and S. Yamamoto for their fruitful 
discussions. We would like to thank all the NRO staff supporting this 
work. S.T. acknowledges a grant from the National Science 
Council of Taiwan (NSC 97-2112-M-001-003-MY2) in support of this 
work.

\clearpage

\clearpage
\begin{deluxetable}{lllcrr} \tablecaption{Observed Molecular Lines\label{tbl-1}}
\tablewidth{0pt}
\tablehead{\colhead{} &\colhead{} &\colhead{Frequency\tablenotemark{a}}
&\colhead{Einstein $A-$Coefficient} & \colhead{$E_{u}$\tablenotemark{b}} & \colhead{$g_{u}$\tablenotemark{c,d}}\\
\colhead{Molecule} & \colhead{Transition} & \colhead{(GHz)} & \colhead{(s$^{\rm -1}$)} &\colhead{(K)} &\colhead{}}
\startdata
HC$^{18}$O$^{+}$  &$J$ = 1--0  &85.162157  & 3.97 $\times$ 10$^{-5}$\tablenotemark{e} &4.1\tablenotemark{a} &3 \\
C$_{3}$H$_{2}$    &$J_{K_{a}K_{c}}$ = 2$_{12}$--1$_{01}$  &85.338906  &2.55 $\times$ 10$^{-5}$\tablenotemark{f} &4.1\tablenotemark{g} &15 \\
HCS$^{+}$         &$J$ = 2--1  &85.347878  &1.00 $\times$ 10$^{-5}$\tablenotemark{h} &6.1\tablenotemark{a} &5 \\
CH$_{3}$CCH       &$J_{K}$ = 5$_{K}$--4$_{K}$ $K$ = 0,1,2,3  &85.457299\tablenotemark{i} &1.86 $\times$ 10$^{-6}$\tablenotemark{d,i}  &12.3\tablenotemark{d,i}  &11\tablenotemark{i}\\
CH$_{3}$OH        &$J_{K}$ = 6$_{-2}$--7$_{-1}$ E  &85.568074  &1.13 $\times$ 10$^{-6}$\tablenotemark{j} &66.8\tablenotemark{j} &13 \\
C$_{3}$H$_{2}$    &$J_{K_{a}K_{c}}$ = 4$_{32}$--4$_{23}$  &85.656418  &1.67 $\times$ 10$^{-5}$\tablenotemark{k} & 26.7\tablenotemark{g} &27 \\
C$_{4}$H          &$N$ = 9--8, $J$ = $\frac{17}{2}$--$\frac{15}{2}$ &85.67257 &2.78 $\times$ 10$^{-6}$\tablenotemark{d} &20.6\tablenotemark{d} &36\\
$^{13}$CO         &$J$ = 1--0  &110.201353  &6.51 $\times$ 10$^{-8}$\tablenotemark{l} &5.3\tablenotemark{a} &3 \\
CH$_{3}$CN        &$J$ = 6$_{K}$--5$_{K}$ $K$ = 0,1,2,3  &110.383522\tablenotemark{i}  &1.09 $\times$ 10$^{-4}$\tablenotemark{d,i} &18.5\tablenotemark{d,i}  &78\tablenotemark{i} \\
\enddata
\tablenotetext{a}{From the Lovas catalog \cite{lov04}.}
\tablenotetext{b}{Upper-state energy level from the ground state of the relevant symmetric state.}
\tablenotetext{c}{Upper-state degeneracy.}
\tablenotetext{d}{From the JPL Molecular catalog \cite{pic98}.}
\tablenotetext{e}{Adopting the dipole moment $\mu$ = 4.07 (Debye) \cite{hae79}.}
\tablenotetext{f}{Adopting $\mu$ = 3.43 (Debye) \cite{kan87}, and
the line strength $S$ = 1.5 \cite{vrt87}.}
\tablenotetext{g}{From Vrtilek et al. (1987).}
\tablenotetext{h}{Adopting $\mu$ = 1.86 (Debye) \cite{win79}.}
\tablenotetext{i}{For $K$ = 0.}
\tablenotetext{j}{From Xu \& Lovas (1997).}
\tablenotetext{k}{Adopting $\mu$ = 3.43 (Debye) \cite{kan87} and $S$ = 1.75 \cite{vrt87}.}
\tablenotetext{l}{Adopting $\mu$ = 0.112 (Debye) \cite{win79}.}
\end{deluxetable}

\clearpage
\begin{deluxetable}{llccccccccccc}
\tabletypesize{\scriptsize}
\rotate
\tablecolumns{13}
\tablewidth{0pc}
\tablecaption{Observed Line Parameters towards L1521F-IRS and IRAM 04191\tablenotemark{a}\label{tbl-2}}
\tablehead{\colhead{} &\colhead{} &\multicolumn{5}{c}{L1521F-IRS} &\colhead{} &\multicolumn{5}{c}{IRAM 04191} \\
\cline{3-7} \cline{9-13}
\colhead{Molecule} &\colhead{Transition} &\colhead{$\int T_{MB} dv$} &\colhead{$T_{MB}$} &\colhead{rms} &\colhead{$\Delta v$} &\colhead{$v_{\rm LSR}$} &
                              \colhead{} &\colhead{$\int T_{MB} dv$} &\colhead{$T_{MB}$} &\colhead{rms} &\colhead{$\Delta v$} &\colhead{$v_{\rm LSR}$} \\
\colhead{} & \colhead{}                  &\colhead{(K km s$^{-1}$)} &\colhead{(K)} &\colhead{(K)} &\colhead{(km s$^{-1}$)} &\colhead{(km s$^{-1}$)} 
                             &\colhead{} &\colhead{(K km s$^{-1}$)} &\colhead{(K)} &\colhead{(K)} &\colhead{(km s$^{-1}$)} &\colhead{(km s$^{-1}$)}}
\startdata
HC$^{18}$O$^{+}$ &$J$ = 1--0 &0.17 &0.46 &0.068 &0.34 &6.15 & &0.32 &0.19 &0.062 &1.64 &6.59 \\
C$_{3}$H$_{2}$   &$J_{K_{a}K_{c}}$ = 2$_{12}$--1$_{01}$\tablenotemark{b} &1.82 &3.87 &0.075 &0.50 &6.40  & &1.37 &2.63 &0.069 &0.40 &6.56 \\
HCS$^{+}$        &$J$ = 2--1 &0.19 &0.51 &0.076 &0.35 &6.42 & &0.09 &0.26 &0.074 &0.33 &6.55 \\
CH$_{3}$CCH      &$J_{K}$ = 5$_{0}$--4$_{0}$            &0.31 &0.80 &0.070 &0.37 &6.39  & &0.16 &0.30 &0.073 &0.51 &6.55 \\
CH$_{3}$CCH      &$J_{K}$ = 5$_{1}$--4$_{1}$            &0.39 &0.90 &0.070 &0.41 &6.38  & &0.07 &0.25 &0.073 &0.28 &6.52 \\
CH$_{3}$CCH      &$J_{K}$ = 5$_{2}$--4$_{2}$            &\nodata  &\nodata  &0.070 &\nodata  &\nodata   & &\nodata  &\nodata  &0.073 &\nodata  &\nodata  \\
CH$_{3}$CCH      &$J_{K}$ = 5$_{3}$--4$_{3}$            &\nodata  &\nodata  &0.070 &\nodata  &\nodata   & &\nodata  &\nodata  &0.073 &\nodata  &\nodata  \\
CH$_{3}$OH       &$J_{K}$ = 6$_{-2}$--7$_{-1}$ E        &\nodata &\nodata &0.067 &\nodata &\nodata & &\nodata &\nodata &0.068 &\nodata &\nodata \\
C$_{3}$H$_{2}$   &$J_{K_{a}K_{c}}$ = 4$_{32}$--4$_{23}$ &\nodata &\nodata &0.060 &\nodata &\nodata & &\nodata &\nodata &0.057 &\nodata &\nodata \\
C$_{4}$H         &$N$ = 9--8, $J$ = $\frac{17}{2}$--$\frac{15}{2}$ &0.35 &0.81 &0.060 &0.41 &6.37 & &0.10 &0.27 &0.057 &0.34 &6.61 \\
$^{13}$CO        &$J$ = 1--0\tablenotemark{b} &6.79 &6.83 &0.097 &1.11 &6.64 & &4.86 &8.00 &0.089 &1.19 &6.77 \\
                 &                            &5.03 &5.57 &0.097 &0.17 &7.37 & & & & &    & \\
CH$_{3}$CN &$J$ = 6$_{K}$--5$_{K}$ &\nodata &\nodata &0.094 &\nodata &\nodata  & &\nodata  &\nodata  &0.093 &\nodata  &\nodata \\
\enddata
\tablenotetext{a}{These parameters are derived from the one-component Gaussian fitting to
the spectra shown in Figure \ref{spec}, unless otherwise noted. Here, $T_{MB}$, $\Delta v$, and $v_{\rm LSR}$
are the peak main-beam brightness temperature,
FWHM width, and the central LSR velocity of the Gaussian. $\int T_{MB} dv$ 
is the integrated value of the Gaussian.}
\tablenotetext{b}{Non-Gaussian spectral shapes. $T_{MB}$, $\Delta v$, $v_{\rm LSR}$, and
$\int T_{MB} dv$ are the peak brightness temperature, FWHM line width, peak velocity,
and the integrated intensity over the detected velocity range.}
\end{deluxetable}

\clearpage
\begin{deluxetable}{lccccc} \tablecaption{Molecular Column Densities and Abundances\label{tbl-3}}
\tablewidth{0pt}
\tablehead{\colhead{} &\multicolumn{2}{c}{Column Density} &\colhead{} &\multicolumn{2}{c}{Abundance}\\
\cline{2-3} \cline{5-6}
\colhead{} &\colhead{L1521F} &\colhead{IRAM 04191} &\colhead{} &\colhead{L1521F} &\colhead{IRAM 04191}\\
\colhead{Molecule} &\multicolumn{2}{c}{$\times$ 10$^{12}$ cm$^{-2}$} &\colhead{} &\multicolumn{2}{c}{$\times$ 10$^{-11}$}}
\startdata
HC$^{18}$O$^{+}$ &0.18 $\pm$ 0.02 &0.35 $\pm$ 0.03 & &0.19 $\pm$ 0.03 &0.27 $\pm$ 0.04 \\
C$_{3}$H$_{2}$   &8.9 $\pm$ 0.9   &6.7 $\pm$ 0.7   & &9.6 $\pm$ 1.4   &5.3 $\pm$ 0.7 \\
HCS$^{+}$        &1.2 $\pm$ 0.1   &0.56 $\pm$ 0.10 & &1.2 $\pm$ 0.2   &0.44 $\pm$ 0.09 \\
CH$_{3}$CCH      &32.9 $\pm$ 3.3  &16.9 $\pm$ 2.1  & &35.6 $\pm$ 5.0  &13.3 $\pm$ 2.1 \\
C$_{4}$H         &87.4 $\pm$ 8.7  &24.6 $\pm$ 3.2  & &94.4 $\pm$ 13.3 &19.3 $\pm$ 3.2 \\
CH$_{3}$CN       &$<$ 0.04\tablenotemark{a} &$<$ 0.04\tablenotemark{a} & &$<$ 0.05\tablenotemark{a} &$<$ 0.03\tablenotemark{a} \\
\enddata
\tablenotetext{a}{1$\sigma$ upper limit.}
\end{deluxetable}

\clearpage
\begin{deluxetable}{lcccccccc} \tablecaption{Evolutional Parameters of Dense Cores\label{tbl-4}}
\tabletypesize{\scriptsize}
\rotate
\tablewidth{0pt}
\tablehead{\colhead{Object} &\colhead{Type} &\colhead{$N$(N$_{2}$H$^{+}$)/$N$(CCS)}
&\colhead{$N$(N$_{2}$D$^{+}$)/$N$(N$_{2}$H$^{+}$)} &\colhead{CO f$_{D}$} &\colhead{$n_{\rm H_2}$}
&\colhead{$N$(CH$_{3}$CCH)} &$N$(C$_{4}$H) &\colhead{ref.\tablenotemark{a}}\\
\colhead{} &\colhead{} &\colhead{}
&\colhead{} &\colhead{} &\colhead{(10$^{5}$ cm$^{-3}$)} &\colhead{(10$^{13}$ cm$^{-2}$)} &\colhead{(10$^{14}$ cm$^{-2}$)} &\colhead{}}
\startdata
L1521B     &starless &0.03 $\pm$ 0.02 &\nodata       &\nodata      &\nodata &\nodata &\nodata &1 \\
L1498      &starless &0.43 $\pm$ 0.06 &0.04 $\pm$ 0.01 &7.5 $\pm$ 2.5  &1.0 $\pm$ 0.7     &\nodata &\nodata &1,2 \\
L1521F     &VeLLO    &0.52 $\pm$ 0.07 &0.10 $\pm$ 0.02 &15.0 $\pm$ 3.6 &11.0 $\pm$ 1.8    &3.3 $\pm$ 0.3    &0.87 $\pm$ 0.09    &2,3,4 \\
L1544      &starless &0.92 $\pm$ 0.16 &0.23 $\pm$ 0.04 &14.0 $\pm$ 3.4 &14.0 $\pm$ 2.2    &\nodata &\nodata &1,2 \\
IRAM 04191 &VeLLO    &1.18 $\pm$ 0.21 &\nodata       &20.0 $\pm$ 2.8\tablenotemark{b}      &10.0 $\pm$ 1.6    &1.7 $\pm$ 0.2 &0.25 $\pm$ 0.03 &1,4,5,6 \\
\enddata
\tablenotetext{a}{References: (1) Aikawa et al. 2003; (2) Crapsi et al. 2005a;
(3) Shinnaga et al. 2004; (4) This work; (5) Belloche et al. 2002;
(6) Belloche \& Andr\'e 2004}
\tablenotetext{b}{Derived from our unpublished C$^{18}$O ($J$=2--1) data taken with the SMT.}
\end{deluxetable}

\end{document}